\documentclass[AMA,STIX1COL]{WileyNJD-v2}
\usepackage{longtable} 
\usepackage{float}
\usepackage{hyperref} 
\hypersetup{                        
    colorlinks=true,                
    breaklinks=true,                
    urlcolor= black,                
    linkcolor= blue,                
    bookmarksopen=false,
    filecolor=black,
    citecolor=blue,
    linkbordercolor=blue
}

\articletype{}
\received{}
\revised{}
\accepted{}

\begin{document}

\title{Decentralized approaches for Autonomous Vehicles Coordination}

\author[LG, GC, MM]{Luca Gherardini$^1$, Giacomo Cabri$^2$, Manuela Montangero$^2$}
\address{\orgname{$^1$ Sano centre for computational personalised medicine, Czarnowiejska 36 building C5, 30-054 Kraków, Poland}}
\address{\orgname{$^2$ Dipartimento di Scienze Fisiche, Informatiche e Matematiche, Università di Modena e Reggio
Emilia, Via Campi 213/B, 41125 Modena, Italy}}
\corres{l.gherardini@sanoscience.org, giacomo.cabri@unimore.it, manuela.montangero@unimore.it}

\authormark{L. Gherardini \textsc{et al.}}

\abstract[Abstract]{The coordination of autonomous vehicles is an open field that is addressed by different researches comprising many different techniques.
In this paper we focus on decentralized approaches able to provide adaptability to different infrastructural and traffic conditions. We formalize an Emergent Behavior Approach that, as per our knowledge, has never been performed for this purpose, and a Decentralized Auction approach. We compare them against existing centralized negotiation approaches based on auctions and we determine under which conditions each approach is preferable to the others. 
} 
\maketitle              

\section{Introduction}
\label{sec:introduction}
Autonomous vehicles are gaining attention from both industrial and research communities, because on the one hand they can change  our mobility~\cite{menon2018,roth2018} while on the other hand they still need research to make it everyday normality~\cite{preparing}.
These new perspectives open for further revolutions, in particular for the coordination of autonomous vehicles, which can improve exploitation of shared resources~\cite{bertogna2017adaptive}, such as intersections and parking slots~\cite{parking}.
One of the most popular conceptualization regards the use of \emph{auctions} to allocate and contend such resources~\cite{AuctionPeterStone, consensus-based_task_allocation}, in many shades.
Different approaches for autonomous vehicle coordination have been proposed, for which we recommend the survey and taxonomy available in~\cite{mariani2021coordination}.

Most of the existing researches focus on the \emph{negotiation} approach~\cite{AuctionPeterStone} in a \emph{centralized} perspective. Conversely, in this paper we focus on two decentralized techniques to coordinate autonomous vehicles  intersection crossing, belonging to \emph{emergent behavior}~\cite{agent-based_simulation_in_emergent_behavior}, showing highly decentralized and dynamic properties, and a decentralized implementation of our version of auction algorithm ~\cite{cabri2021auction} at \emph{negotiation} level.
The term emergent behavior denotes a system made of many elements with a relatively simple set of rules, interacting at low level to build an emerging higher structures with more sophisticated characteristics. Complexity arises from interactions, not from individual agents, making the succession of actions changeable and not trivial to follow or to predict. It is important to point out the unawareness of the elements about their role in the whole picture.
The {negotiation} between vehicles aims to determine the best solution to a collective problem (i.e. reducing commuting time) with each element deliberately participating in the process, resulting in a well predictable course of action. 
We compare them with two centralized approaches already existing in the literature~\cite{cabri2021auction} exploiting auctions at negotiation level. These two systems mainly differed for the rules through which determining the winner of the auction, renamed \emph{cooperative} and \emph{competitive}. The former allowed all vehicles to cross the intersection following the order of their bids; the latter allowed only the winner, leaving the defeated vehicles to participate in the next auction.

We simulated these approaches 
and collected data (e.g., waiting times) useful for a detailed comparison accounting for traffic volumes and parameters. The simulations we performed have been described in section ~\ref{sec:methods}, displaying and analyzing the obtained results in section~\ref{sec:results} for the metrics adopted, and lastly we discuss them in section~\ref{sec:conclusions}.

\section{Methods}
\label{sec:methods}
We implemented and tested all strategies for intersection management by using SUMO (Simulator of Urban MObility - \url{https://sumo.dlr.de}), a simulator allowing a fine granularity in vehicles management. By exploiting the SUMO API called ``TraCI'' (Traffic Control Interface - \url{https://sumo.dlr.de/docs/TraCI.html}), written in Python, we were able to implement specific vehicle behaviors and to monitor and to collect vehicles waiting times. We provide the implemented algorithms and benchmarks in a dedicated GitHub repository (\url{https://github.com/LucaGherardini/DecentralizedCoordinationAlgorithms}).

We tested the emergent behavior approach under different initial configurations and we compared its performances against centralized~\cite{cabri2021auction} and decentralized auction-based approaches, whose details are given later in this section. 

We defined a virtual scenario consisting of a $5\times5$ Manhattan style map and, for each approach and each parameter choice, we run 10 tests with 100 vehicles (Vehicles Spawned ({\bf VS})), commuting in random routes ({\bf Rts}) for 10000 simulation time steps ({\bf Stp}), each step corresponding to one second. 
A random route is sampled when a vehicle is traveling to the end of its last assigned edge (or road), picking with uniform probability among the edges linked to its end. The process is reiterated considered the feasible edges linked to the end of the last picked one, until a route of the same length of the original one (12 edges) is formed. Consider that the first route of a vehicle is predetermined, and the random sampling is performed after its completion.

In the following subsections we presents the different approaches simulated in our work.

\subsection{Emergent Behavior}
The \emph{Emergent Behavior} ({\bf EB}) defines an \emph{ecosystem} in which vehicles interact through reciprocal observations without direct one-to-one communication. Each vehicle uses its perception and perspective of the surrounding world to make the best choices to reach its prefixed goal. 
Each one has an internal state representing its ``hurry'', defined by means of its experienced waiting time: the vehicle hurry increases and decreases, respectively, during slow and free motion periods.
Vehicles expose their internal state to the others nearby, giving to the one with the highest ``hurry'' the right to cross the intersection.

In details, ``hurry'' ({\bf H} in \tablename~\ref{if-df}) is always a non-negative value which increases and decreases independently for each vehicle at the end of each simulation time step by applying an Increasing Function ({\bf IF}) and a Decreasing Function ({\bf DF}) by means of Increasing and Decreasing Coefficients (resp. {\bf IC} and {\bf DC}). \tablename~\ref{if-df} reports IFs and DFs used in the simulations.

\renewcommand{\arraystretch}{1.2} 
\begin{longtable}{p{3cm}|c|p{9cm}}
    \label{if-df}
    {\bf Function (IF/DF)} & {\bf Mathematical Definition} & {\bf Description} \\ \hline
    Linear (Lin) & $H_{t+1} = H_t + C$ & Linear update at fixed rate $C$ \\ \hline
    Logarithmic (Log) & $H_{t+1} = H_t + C * ln(H_t + 2)$ & Coefficient $C$ regulates the strength and the polarity of the change caused by the logarithm. The argument of the latter is incremented by 2 to always have positive values. \\ \hline
    Grower (Gro) & $H_{t+1} = H_t + \max \left\{ \frac{H_t * C}{100}, C \right\}$ & $C$ is both a lower bound for the  strength update and a measure of the percentage of the hurry used for increasing/decreasing it. \\ 
    \caption{Description of functions used as Increasing and Decreasing Functions (IF and DF) for the emergent approach. Coefficient C is positive if the vehicle is waiting or negative if in free motion.} 
\end{longtable}

Moreover, vehicles in the same lane and within a Spreading Range  ({\bf SR}) will interact to influence the reciprocal ``hurry''. When a group of adjacent vehicles converges to the same value, they are saying to organize into a ``platoon'', i.e., a column formation of vehicles with similar ``hurry'' that will be allowed to cross the intersection all together. This phenomenon is not explicitly modeled as a property of the system, but it's a macro characteristic ``emerging'' (from here is the name of the approach) from microscopic interactions (vehicle-to-vehicle). \tablename~\ref{spread} shows Spreading functions used in the simulations to compute the interactions between a vehicle $v$ and all of its neighbors $N_v$.   
The Distance ({\bf D}) between the vehicles damps the force of this interaction in a measure depending on the Distance Magnitude ({\bf DM}).

\begin{table}[t]
\begin{longtable}{p{3cm}|c|p{9cm}}
    \label{spread}
    Spreading Function & Mathematical Definition & Description \\ \hline
    Standard (Std) & $\sum_n^{N_v}(|H_n - H_v| * \frac{DM}{D_{nv}})$ & $H_n$ and $H_v$ are, respectively, the neighbor's hurry and the vehicle hurry. \\ \hline
    Distance-Based Logarithmic (DBL) & $\sum_n^{N_v}(ln(|H_n - H_v| + 1) * \frac{DM}{D_{nv}})$ & Variant considering a softer influence by vehicles nearby. \\ \hline
    Range-Based Logarithmic (RBL) & $\sum_n^{N_v}(ln(|H_n - H_v| + 1) * \frac{SR * DM}{D_{nv}})$ & This variant considers the ratio between Spreading Range and the actual Distance with the current neighboring vehicle. \\
    
    \caption{Spreading functions used in simulations, $\forall n \in N_v$}
\end{longtable}
\end{table}
\renewcommand{\arraystretch}{1} 

\subsection{Centralized Auction-based Approaches} 
 Inspired by the concept of economic auction, in this approach vehicles have a trip budget at their disposal and use it to place bids and cross intersections. Auctions are handled by city infrastructure at crossing sites and might allow all participants (\emph{Cooperative} approach ({\bf Coop})) or just the winner (\emph{Competitive} approach ({\bf Comp})) to cross the intersection. 
 
 In the former case, the auction aims at defining the order in which the participants cross the intersections, in the latter case the auction is exploited to define the only participant that crosses the intersection.
 
 Both alternatives are characterized by the parameters described in \tablename~\ref{auction-tab}. 

\begin{longtable}{c|p{16cm}}
    \label{auction-tab} 
    {\bf CP} & \emph{Crossing Policy} determines which bids are charged:  Only-Winner-Pays ({\bf OWP}) charges only the winner, All-Vehicles-Pay ({\bf AVP}) charges all vehicles. \\ \hline
    {\bf MCA} & \emph{Minimum Cars to Auction} defines the minimum number of vehicles that must be close to a intersection in order to start an auction (default is set to {\bf 2}). \\ \hline 
    {\bf E} & \emph{Enhancement} is a mechanism to boost bids coming from overcrowded lanes, to avoid starvation, and depends logarithmically on the number of vehicles in lanes. It can be activated or not ({\bf Y/N}). \\ \hline 
    {\bf Bdn} &  \emph{Bidding} is the bidding strategy adopted by vehicles: Balanced (B), i.e., a fraction of the budget is used depending on how many intersections are to be crossed; Random ({\bf R}) 
    with uniform probability in the budget interval. \\ \hline
    
    {\bf Spn}  & \emph{Sponsorship} (only in Competitive approach) is the vehicle budget percentage spendable to help the front line vehicle win the auction ({\bf 25, 50, 75\%}). \\ \hline
    {\bf Rts} & \emph{Routes} defines if reassigning cyclically the same route to a vehicle (S) or computing a random one each time a vehicle completes it ({\bf R}). \\
    \caption{Parameters considered for Auction Strategies. Values used in simulations are given in bold.}
\end{longtable} 

\subsection{Decentralized Auction-based approaches} 
We implemented a novel decentralized version of the auction-based approach with no need of dedicated auctioneers. The lack of a central entity causes this approach not to be a direct translation of centralized protocols as, to keep the distributed protocol efficient and applicable in a real context, it does not support all the parameters as the latter does. Therefore, more complex mechanisms, such as enhancement and sponsorship, are not implemented. 

In this approach, a vehicle close to an intersection
broadcasts its bid to other vehicles near the intersection to check their ranking in the auction.
Vehicles then access the intersection according to their bids (higher first), allowing newcomers with higher bids to overtake. This implementation can be intended as a middle point between cooperative and competitive protocols, as if no vehicle with a competitive bid arrives at the intersection, then participants would cross in a cooperative fashion. Conversely, if the newcomer has a bid higher than some other participants, then it cross before then, without causing the auction to repeat.

\pagebreak
\section{Results}
\label{sec:results}

\begin{figure}
    \centering
    \includegraphics[width=.7\textwidth]{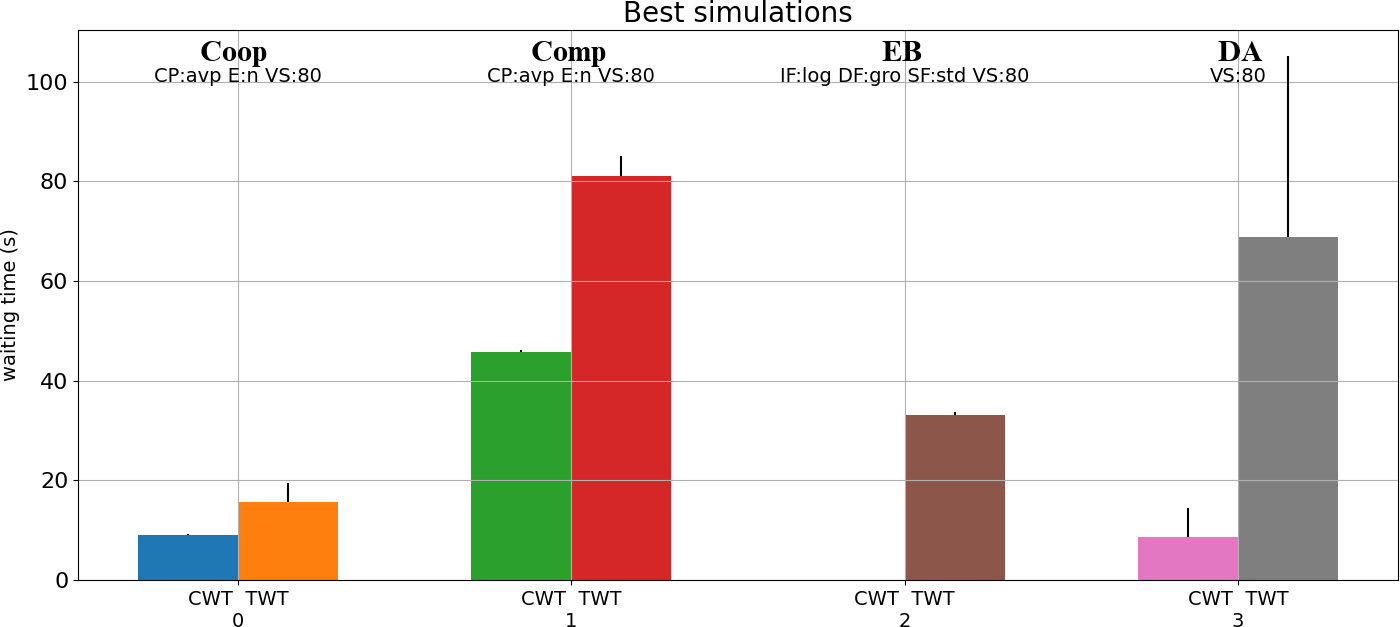}
    \caption{Comparison of the best simulations for both CWT and TWT, as highlighted in tables \ref{sims-coop}, \ref{sims-da} \ref{sims-comp}, and \ref{sims-eb} for 80 vehicles.}
    \label{fig:best_comp}
\end{figure}

To test the different approaches, we used random routing (and bidding for auction-based approaches) in all simulations. 
Thorough simulations we measured waiting times experienced by vehicles: \emph{Crossing Waiting Time} (CWT) is the average time needed to cross intersection, starting when the vehicle is the first of its lane; \emph{Traffic Waiting Time} (TWT) is the average waiting time spent in line among other vehicles in slow/no motion. {\tablename}s \ref{sims-coop}, \ref{sims-da} \ref{sims-comp}, and \ref{sims-eb} show the best performing configurations for each approach with the corresponding CWT/TWT (in seconds) when varying the parameters values. Values in bold highlight the lowest (best) total waiting times (CWT+TWT) for each approach and for each volume of traffic. Figure~\ref{fig:best_comp} compares the different approaches showing, for each one, the best performing configuration.

\begin{table}
\parbox{.60\linewidth}{
\centering
\begin{tabular}{c c | c c | c c}
     \multicolumn{2}{c}{{\bf Coop}} & \multicolumn{2}{c}{E=N} & \multicolumn{2}{c}{E=Y}\\ \hline
    {\bf CP} & {\bf VS} & {\bf CWT ($\pm \sigma$)} & {\bf TWT ($\pm \sigma$)} & {\bf CWT ($\pm \sigma$)} & {\bf TWT ($\pm \sigma$)} \\ \hline 
    \multirow{3}{*}{AVP} 
                & {\bf 80} & {\bf 9.0 $\pm$0.3} & {\bf 15.6 $\pm$3.9} & 
                10.3 $\pm$ 0.7 & 20.3 $\pm$ 5.1 \\
    
                & 100 & 10.2 $\pm$0.4 & 20.7 $\pm$4.6 & 
                12.3 $\pm$ 0.8 & 31.2 $\pm$ 6.7 \\
    
                & {\bf 120} & {\bf 12.9 $\pm$1.6} & {\bf 19.3 $\pm$ 5.2} &
                13.6 $\pm$ 0.5 & 33.9 $\pm$ 7.7 \\ \hline 
    
    \multirow{3}{*}{OWP} 
                & 80 & 9.1 $\pm$ 0.4 & 17.0 $\pm$4.8 & 
                10.3 $\pm$0.7 & 21.9 $\pm$ 5.0 \\
    
                & {\bf 100} & {\bf 10.7 $\pm$ 0.8} & {\bf 19.2 $\pm$ 5.7} & 
                12.5 $\pm$ 0.5 & 29.2 $\pm$ 7.0 \\
    
                & 120 & 12.0 $\pm$ 0.0 & 26.3 $\pm$ 5.2 & 
                13.8 $\pm$ 0.4 & 32.3 $\pm$8.0 \\
    \end{tabular}
    \caption{Cooperative approach simulations for \emph{MCA}=2; \emph{Bdn}=R; \emph{Rts}=R; \emph{Stp}=10'000.}
    \label{sims-coop}
}
\hfill
\parbox{.30\linewidth}{
\centering
\begin{tabular}{c c c}
    \multicolumn{3}{c}{{\bf Decentralized Auction}}\\ \hline
    {\bf VS} & {\bf CWT ($\pm \sigma$)} & {\bf TWT ($\pm \sigma$)} \\ \hline 
    80 & 8.7 $\pm$ 5.8 & 68.9 $\pm$ 36.2 \\ \hline
    100 &  6.2 $\pm$ 3.4 & 72.1 $\pm$ 31.6 \\ \hline
    120 & 5.8 $\pm$ 2.1 & 77.6 $\pm$ 14.5 \\ \hline
    
\end{tabular} 
\caption{Configurations inherent to the Decentralized Auction approach.}
\label{sims-da} 
}
\end{table}

\begin{longtable}{c c | c c | c c}
\multicolumn{2}{c}{{\bf Comp}} & \multicolumn{2}{c}{E=N} & \multicolumn{2}{c}{E=Y}\\ \hline
    {\bf CP} & {\bf VS} & {\bf CWT ($\pm \sigma$)} & {\bf TWT ($\pm \sigma$)} & {\bf CWT ($\pm \sigma$)} & {\bf TWT ($\pm \sigma$)} \\ \hline
    \multirow{3}{*}{OWP}
                    & 80 & 45.9 $\pm$ 0.5 & 82.5 $\pm$ 2.7     & 
                    48.1 $\pm$ 0.7 & 91.1 $\pm$ 4.1 \\
    
                    & 100 & 49.6 $\pm$ 0.7 & 115.4 $\pm$ 4.4   & 
                    51.0 $\pm$ 0.6 & 122.1 $\pm$ 7.8 \\
                    
                    & 120 & 52.3 $\pm$ 0.6 & 150.2 $\pm$ 7.2   & 
                    54.7 $\pm$ 0.6 & 159.1 $\pm$ 9.8 \\
    \hline
    \multirow{3}{*}{AVP} 
                    & {\bf 80} & {\bf 45.7 $\pm$ 0.5} & {\bf 81.0 $\pm$ 4.1} & 
                    48.4 $\pm$ 0.7 & 90.1 $\pm$ 2.3 \\
                    & {\bf 100} & {\bf 49.4 $\pm$ 0.8} & {\bf 114.2 $\pm$ 5.9}   & 
                    51.3 $\pm$ 0.8 & 124.2 $\pm$ 5.9 \\
                    & {\bf 120} & {\bf 52.8 $\pm$ 0.6} & {\bf 149.9 $\pm$ 5.0}   & 
                    55.0 $\pm$ 0.9 & 163.0 $\pm$ 3.8 \\
    \caption{Competitive approach simulations for \emph{MCA}=2; \emph{Bdn}=R; \emph{Rts}=R; \emph{Stp}=10'000; \emph{Spn}=25.}
    \label{sims-comp}
\end{longtable} 

\begin{longtable}{c c c | c c | c c | c c}
    \multicolumn{3}{c}{{\bf EB}} & \multicolumn{2}{c}{{\bf RBL}} & \multicolumn{2}{c}{{\bf DBL}} & \multicolumn{2}{c}{{\bf Std}}\\ \hline
    {\bf IF} & {\bf DF} & {\bf VS} & {\bf CWT ($\pm \sigma$)} & {\bf TWT ($\pm \sigma$)} & {\bf CWT ($\pm \sigma$)} & {\bf TWT ($\pm \sigma$)} & {\bf CWT ($\pm \sigma$)} & {\bf TWT ($\pm \sigma$)} \\ \hline
        Lin & Gro & 80 
        & 0.0 $\pm$ 0.0 & 36.3 $\pm$ 0.5 
        & 0.0 $\pm$ 0.0 & 37.5 $\pm$ 3.3 
        & 0.0 $\pm$ 0.0 & 33.3 $\pm$ 0.6 \\
        
        Log & Log & 80 
        & 0.0 $\pm$ 0.0 & 39.4 $\pm$ 2.2
        & 0.1 $\pm$ 0.3 & 105.6 $\pm$ 13.0
        & 0.0 $\pm$ 0.0 & 35.2 $\pm$ 1.3 \\
        
        Lin & Lin & 80 
        & 0.0 $\pm$ 0.0 & 52.2 $\pm$ 4.8 
        & 1.3 $\pm$ 0.5 & 210.5 $\pm$ 50.4
        & 0.0 $\pm$ 0.0 & 34.6 $\pm$ 0.9 \\

        Log & Gro & 80 
        & 0.0 $\pm$ 0.0 & 36.4 $\pm$ 0.8
        & 0.0 $\pm$ 0.0 & 36.4 $\pm$ 1.4
        & {\bf 0.0 $\pm$ 0.0} & {\bf 33.2 $\pm$ 0.6} \\
        
        \hline
        
        Lin & Gro & 100 
        & 0.0 $\pm$ 0.0 & 40.9 $\pm$ 0.7  
        & 0.0 $\pm$ 0.0 & 45.4 $\pm$ 2.8
        & 0.0 $\pm$ 0.0 & 38.8 $\pm$ 0.4 \\
        
        Log & Log & 100 
        & 0.0 $\pm$ 0.0 & 42.0 $\pm$ 1.3
        & 1.0 $\pm$ 0.0 & 113.6 $\pm$ 15.3   
        & 0.0 $\pm$ 0.0 & 39.2 $\pm$ 0.6 \\
        
        Lin & Lin & 100 
        & 0.0 $\pm$ 0.0 & 50.8 $\pm$ 2.9
        & 2.2 $\pm$ 2.00 & 189.8 $\pm$ 52.3
        & 0.0 $\pm$ 0.0 & 39.2 $\pm$ 0.6   \\ 
        
        Log & Gro & 100
        & 0.0 $\pm$ 0.0 & 41.2 $\pm$ 0.6
        & 0.0 $\pm$ 0.0 & 44.2 $\pm$ 1.4
        & {\bf 0.0 $\pm$ 0.0} & {\bf 38.6 $\pm$ 0.9} \\
        
        \hline
        
        Lin & Gro & 120 
        & 0.0 $\pm$ 0.0 & 47.9 $\pm$ 0.7
        & 0.0 $\pm$ 0.0 & 53.6 $\pm$ 1.5
        & 0.1 $\pm$ 0.3 & 47.5 $\pm$ 1.5 \\
        
        Log & Log & 120 
        & 0.0 $\pm$ 0.0 & 48.4 $\pm$ 1.2
        & 1.0 $\pm$ 0.0 & 134.5 $\pm$ 29.9
        & 0.0 $\pm$ 0.0 & 46.6 $\pm$ 1.2 \\
      
        Lin & Lin & 120
        & 0.0 $\pm$ 0.0 & 57.9 $\pm$ 4.7
        & 2.5 $\pm$ 0.7 & 257.0 $\pm$ 50.1
        & {\bf 0.0 $\pm$ 0.0} & {\bf 45.7 $\pm$ 1.1} \\       
        Log & Gro & 120
        & 0.0 $\pm$ 0.0 &  49.2 $\pm$ 1.5
        & 0.0 $\pm$ 0.0 & 55.4 $\pm$ 3.7
        & 0.1 $\pm$ 0.3 & 46.9 $\pm$ 1.8 \\
    \caption{Emergent approach simulations for \emph{IC}=10; \emph{DC}=10; \emph{SR}=100; \emph{DM}=10; \emph{SP}=AN.}
    \label{sims-eb} 
\end{longtable} 

\pagebreak
\subsection{Discussion}
Our simulations tested the effectiveness of some intersection management approaches, and showed that the centralized Cooperative approach has the best average waiting time.
These results allowed us to select the most performing configurations for each approach and for each volume of traffic, then we used a direct comparison to determine the best.
The Cooperative auction based approach neatly showed the best performance, in particular in the "AVP-N" combination, shadowing the Competitive auction based approaches on every configuration. Its interesting to see the dichotomy derived by the simple difference in the handling of the outcome of the auction. The former is actually pursuing a \emph{fair} approach avoiding to  excessively penalize vehicles defeated in auctions, and for this it resolves well balanced waiting times, while the second and more rigid approach causes higher queues and consequently stumbles in prolonged waiting.

The Emergent Behavior approach showed a qualitatively comparable performance, especially in the ``logarithm-grower-std'' combination, with a competitive average time and a low standard deviation. 

Finally, Decentralized Auction shows low CWT in contrast to high and fitful TWT. Interestingly, these results can be interpreted to identify strong suits and weak points of every model, in particular with respect to traffic volume, to identify solutions apt to secondary or rural intersections, and to infrastructural requirements.

From our experiments it emerges that the adoption of an approach depends on the conditions of the intersection(s) and on the policies that the local administration aims at applying.
For instance, intersections with low traffic can benefit from approaches that do not require an overhead time to be applied. Another example can be a place with a high degree of pollution, where the local authority applies an approach that reduces the long queues at the expense of faster (or richer) drivers.

Another important aspects that must be taken into account is the availability of drivers' (virtual) money to pay in auction based approaches. If a local administration aims at gaining money from the intersections and traffic management, then this approach shall be   preferred.

\section{Conclusions}
\label{sec:conclusions}

Cooperative and Competitive Auction approaches have the merit of constituting an intuitive and adaptable scenario for the transition phase toward autonomous driving vehicles, being able to handle intersections even in presence of human driven vehicles. On the other hand, their infrastructure requirements could be an initial hindrance due to the implementation costs bounded to them and represent a single point of failure for the coordination process. 

Emergent Behavior algorithms show performance comparable with the Cooperative approach, including configuration with an highly stable trend during trials. Furthermore, they embodies scalability and independence from external coordinating structures, at the cost of high requirements on vehicles themselves, which have great responsibilities.  

Decentralized Auction can represent an alternative to Centralized Auction models in absence of human driven vehicles, or in contexts of additional hardware to handle the auctions and notify drivers. This completely decentralized scenario however needs high security requirements able to detect malicious agents. In a recent paper~\cite{delmonte2020scaling} it has been conceptualized a general framework able to scale block-chain security protocol in highly crowded and decentralized contexts, without trade offs on the security side. This approach could find a useful application in this context. Its high TWT suggests their suitability for low volume traffic, such as country intersections. 

With regard to future work, our results open some interesting directions.
First, they point to the necessity of a measure representing the trade-off between feasibility (or complexity) and efficiency to assure practical and incremental approaches toward the total adoption of autonomous driving vehicles.
Second, we have made experiments applying a single approach to all intersections, but it can be interesting to study the results in a heterogeneous network, in particular to understand what happens when different approaches are applied in different intersections or in different clusters of intersections in the same smart city.

A third direction that is worth exploring is how to dynamically switch approaches depending on different traffic conditions, and how to handle the long expected transition through the full adoption of self-driving vehicles.
Finally, we plan to further evaluate and develop parameters we fixed in this comparison.

\medskip
\noindent {\bf Acknowledgments. }
This publication is supported by the European Union’s Horizon 2020 research and
innovation programme under grant agreement Sano No 857533, and by Sano project carried out within the International
Research Agendas programme of the Foundation for Polish Science, co-financed by
the European Union under the European Regional Development Fund.

%
%
%
 \bibliography{main}
\end{document}